\documentclass[10.5pt, lettersize,journal]{IEEEtran}
\usepackage{amsmath,amsfonts}
\usepackage{algorithmic}
\usepackage{algorithm}
\usepackage{array}
\usepackage{makecell}
\usepackage[caption=true,font=footnotesize,labelfont=rm,textfont=rm]{subfig}
\usepackage{textcomp}
\usepackage {color}
\usepackage{stfloats}
\usepackage{url}
\usepackage{verbatim}
\usepackage{graphicx}
\usepackage{cite}
\usepackage{balance}
\usepackage{booktabs}
\usepackage{multirow}
\usepackage{soul}
\usepackage[table]{xcolor} 

\usepackage{dashrule}

\title{Smark: A Watermark for Text-to-Speech Diffusion Models via Discrete Wavelet Transform\\[0.3cm]
\author{Yichuan Zhang, Chengxin Li, Yujie Gu}
\thanks{Faculty of Information Science and Electrical Engineering, Kyushu University, Fukuoka, Japan. (Emails: zhang.yichuan.307@s.kyushu-u.ac.jp, l12chengxin@gmail.com, guyujie2016@gmail.com)}
}
\begin{document}

\maketitle

\begin{abstract}

Text-to-Speech (TTS) diffusion models generate high-quality speech, which raises challenges for the model intellectual property protection and speech tracing for legal use. Audio watermarking is a promising solution. However, due to the structural differences among various TTS diffusion models, existing watermarking methods are often designed for a specific model and degrade audio quality, which limits their practical applicability.
To address this dilemma, this paper proposes a universal watermarking scheme for TTS diffusion models, termed Smark. 
This is achieved by designing a lightweight watermark embedding framework that operates in the common reverse diffusion paradigm shared by all TTS diffusion models. 
To mitigate the impact on audio quality, Smark utilizes the discrete wavelet transform (DWT) to embed watermarks into the relatively stable low-frequency regions of the audio, which ensures seamless watermark-audio integration and is resistant to removal during the reverse diffusion process. 
In addition, a statistical analysis using hypothesis testing is applied to verify the reliability of watermark detection.
Extensive experiments are conducted to evaluate the audio quality and watermark performance in various simulated real-world attack scenarios. The experimental results show that Smark achieves superior performance in both audio quality and watermark extraction accuracy. 
Furthermore, ablation studies are conducted to evaluate the necessity and effectiveness of each component of Smark's design, including the use of DWT, watermark embedding sub-bands and timesteps. The results show that embedding watermarks in low-frequency sub-bands in the reverse diffusion process achieves a favorable balance between audio quality and watermark extraction robustness.
\end{abstract}

\begin{IEEEkeywords}Text-to-Speech, Discrete Wavelet Transform, Watermark, Copyright protection\end{IEEEkeywords}

\section{Introduction}

Recent advances in AI have propelled Text-to-Speech (TTS) diffusion models\cite{chen2023lightgrad, huang2022prodiff,chen2020wavegrad, popov2021grad} to the forefront of synthetic speech generation. 
However, their high quality introduces serious risks such as voice deepfakes and copyright infringement. Audio watermarking\cite{chen2023wavmark,pavlovic2022robust, liu2024groot, zhou2024traceablespeech} provides a countermeasure by embedding imperceptible identifiers into synthesized speech, enabling model intellectual property protection and audio source tracing. 

Existing audio watermarking methods can be categorized into two types: post-processing watermarking and generative watermarking. Post-processing methods, such as WavMark~\cite{chen2023wavmark}, embed watermarks into audio after it is generated.
However, post-processing watermarking suffers from inherent drawbacks: it disrupts the original audio's perceptual quality more easily due to the secondary modification, and its robustness is limited against common audio processing attacks.  WavMark~\cite{chen2023wavmark} exhibits significant performance degradation when encountering audio compression or noise addition; another typical post-processing method, Audioseal~\cite{roman2024proactive}, struggles with low extraction accuracy after audio resampling, as the post-embedded watermark is highly susceptible to signal distortion caused by such operations. In contrast, generative watermarking, such as GROOT~\cite{liu2024groot} and TraceableSpeech~\cite{zhou2024traceablespeech}, integrates watermarks during the audio generation process itself, allowing for more seamless embedding and higher output audio quality.
However, existing generative watermarking methods are often degrade the audio quality a lot and model-specific, limiting their broader applicability. 
It is therefore highly desirable to develop a robust and model-agnostic watermarking approach for TTS diffusion models.

In this paper, we propose Smark, a universal generative
watermarking framework for TTS diffusion models. 
Smark introduces operating within the shared reverse diffusion paradigm of all TTS diffusion models, achieving broad compatibility independent of model-specific structures.
To preserve audio quality,
Smark incorporates the discrete wavelet transform (DWT)~\cite{davis1980comparison}
to embed watermarks into perceptually stable low-frequency
components, directly and effectively tackling the key challenges of 
imperceptibility, robustness, and cross-model compatibility in
diffusion-based TTS watermarking. 
Extensive experiments show that Smark outperforms existing methods in both watermark extraction accuracy and
audio quality, indicating superior effectiveness and robustness against
various real-world attacks.

To validate the necessity of each design choice in Smark, we conduct systematic ablation studies, specifically evaluating the impact of using DWT, the selection of DWT sub-bands for watermark embedding, and the timestep for embedding watermarks during the reverse diffusion process. The results confirm that each component is indispensable for ensuring imperceptibility, robustness, and universality; removing or altering any part leads to significant degradation in one or more key performance metrics.

In summary, Smark delivers three key advantages:

1) \textit{Universality}. Smark is applicable to all TTS diffusion models, as it operates within the shared reverse diffusion paradigm, without relying on model-specific architectures.

2) \textit{A superior balance between imperceptibility and robustness}. 
Smark achieves superior performance in both audio quality and watermark accuracy under both benign and attack scenarios by embedding watermarks into perceptually stable low-frequency components via DWT in the reverse diffusion process of TTS diffusion models.

3) \textit{Interpretability}. 
Systematic ablation studies validate the necessity and effectiveness of each Smark design component.

The remainder of this paper is organized as follows. Section \ref{sec:related work} reviews the related work. Section \ref{sec:Preliminaries} recaps the preliminaries. Section \ref{sec:proposed method} presents the proposed method. Section \ref{sec:experiments} describes experiments and results. Section \ref{sec:ablation_study} provides ablation studies validating key design choices. Section \ref{sec:conclusion} concludes the paper.

\begin{table*}[t]
\renewcommand{\arraystretch}{1.35}
\caption{Summary of state-of-the-art audio watermarking schemes for TTS models}
\label{tab:audio_watermarking_schemes}
\centering
\resizebox{\linewidth}{!}{
\begin{tabular}{c l l p{5cm} p{6cm}}
\toprule
\textbf{Scheme Types} & \textbf{Methods} & \textbf{Applicable Objects} & \textbf{Key Advantages} & \textbf{Core Limitations / Requirements} \\
\midrule
\multirow{4}{*}{Post-processing}
& WavMark~\cite{chen2023wavmark} & General audio & Low deployment cost; Model-agnostic & Vulnerable to common audio processing attacks \\
& TimbreWM~\cite{liu2023detecting} & General audio  & Enables source tracing via timbre features &  Limited to timbre-related authentication \\
&
WAtermArk~\cite{guo2025audio} & Voice datasets &  Harmless to audio quality & Dependent on dataset distribution \\
& AudioSeal~\cite{roman2024proactive} & Voice data & Robust to fine-tuning attacks & High computational cost for localization\\
\midrule
\multirow{3}{*}{Generative} 
& GROOT~\cite{liu2024groot} & TTS diffusion models & Simplified embedding via initial Mel & Model-specific; ignores text-semantic alignment \\
& TraceableSpeech~\cite{zhou2024traceablespeech} & Vall-E(neural codec) & Better audio quality retention for Vall-E &  No compatibility with diffusion models \\
 
& \textbf{Smark (Ours)} & TTS diffusion models & Universal compatibility; good audio quality & Requires integration with reverse diffusion process \\
\bottomrule
\end{tabular}
}
\end{table*}

\section{Related Work}
\label{sec:related work}
\subsection{Text-to-Speech  Models}

Traditional TTS models, such as WaveNet~\cite{van2016wavenet}, Tacotron \cite{wang2017tacotron}, and DeepVoice 3~\cite{ping2017deep}, typically generate speech through deterministic one-step mapping, which leads to spectral over-smoothing and loss of fine acoustic details. This fundamental limitation impedes their ability to effectively balance naturalness and generalization, often resulting in perceptually mechanical outputs and prosodic expression.

The emergence of TTS diffusion models has broken this dilemma, and their core advantage lies in achieving speech generation with high naturalness and high controllability through probabilistic diffusion and iterative denoising mechanisms. TTS diffusion models take random noise as the initial input, gradually optimizes the signal during the reverse diffusion process, and finally generates speech with human auditory characteristics, which can accurately match the prosody, timbre, and emotional requirements of text semantics. As a representative model in this field, GradTTS~\cite{popov2021grad} innovatively models the Mel spectrogram (a time-frequency representation aligned with the human auditory system, which can effectively map acoustic signals and subjective auditory perception~\cite{davis1980comparison}) as a Gaussian diffusion process, and introduces text-driven linguistic features and prosodic features to guide the denoising process. This not only ensures the semantic consistency of text-speech,  but also supports fine-grained control over parameters such as speech speed and pitch.

Later  WaveGrad~\cite{chen2020wavegrad} further optimized the generation efficiency by simplifying the waveform generation process through a gradient sampling strategy, shortening the inference time while ensuring sound quality. 
However, the high-quality generation capability of TTS diffusion models also brings challenges to intellectual property protection and content tracing: the high similarity between generated speech and real human speech may be abused, leading to model infringement and audio piracy.
Therefore, designing an effective  universal watermarking scheme that does not harm speech quality is highly desirable for practical applications. 

\subsection{Audio Watermarking}

Audio watermarking provides an effective means for copyright protection and audio authentication. 
Early traditional audio watermarking methods, such as Cox's spread-spectrum scheme~\cite{cox1997secure} and Boney’s psychoacoustic-based approach~\cite{boney1996digital}, were developed prior to deep learning and typically emphasize either robustness at the expense of audible artifacts or imperceptibility with limited embedding capacity, while lacking data-driven adaptive modeling.

The advent of deep learning offers a promising pathway to address these limitations.
By leveraging the learning capabilities of neural networks, deep learning-based watermarking approaches have achieved significant performance improvements. As shown in Table~\ref{tab:audio_watermarking_schemes}, these approaches are mainly divided into two categories:

\subsubsection{Post-processing watermarking}
A watermark is embedded into the already generated audio signals. Typical methods include WavMark~\cite{chen2023wavmark}, TimbreWM~\cite{liu2023detecting}, WatermArk~\cite{guo2025audio}, AudioSeal~\cite{roman2024proactive}, and more. 
WavMark realizes watermark embedding through fine-grained modulation in the waveform domain, does not rely on specific generation models, and has low deployment costs. TimbreWM focuses on the watermark association of speech timbre features, which can realize tracing while retaining the integrity of timbre. WatermArk is designed for protecting voice datasets and claims to be harmless to audio quality, yet its effectiveness is dependent on the dataset distribution. AudioSeal employs localized watermarking to proactively detect voice cloning and shows robustness against fine-tuning attacks, albeit at a higher computational cost for localization tasks. However, the core limitation of this type of method lies in the separability of ``generation-embedding", where watermark embedding requires secondary modification of the generated speech signal, which may not only destroy the natural spectral structure of the speech but also be easily removed during subsequent signal processing.

\subsubsection{Generative watermarking}
A watermark is embedded during the audio generation process of TTS models to realize ``watermarking while generating". Representative works include GROOT~\cite{liu2024groot} and TraceableSpeech~\cite{zhou2024traceablespeech}. GROOT realizes watermark embedding by modifying the initial Mel spectrogram before reverse diffusion. Although the process is simplified, it does not consider the semantic association between text features and the spectrogram, which easily leads to conflicts between the watermark and speech content, resulting in significant sound quality degradation.
TraceableSpeech~\cite{zhou2024traceablespeech} relies on the discrete acoustic token architecture of VALL-E~\cite{wang2023neural} and embeds watermarks through token modulation. However, as it is not designed for a diffusion architecture, it falls outside the scope of this review focusing on diffusion-based TTS models.
So far, existing generative watermarking methods still face the triangular dilemma of ``sound quality - robustness - versatility". And it is highly desired to develop a universal watermarking approach that can adapt to various TTS diffusion models without damaging speech quality.

In this paper, the proposed Smark provides a robust and effective model-agnostic watermarking method for TTS diffusion models. Unlike methods that modify acoustic tokens or initial conditions, Smark leverages the multi-resolution analysis capability of DWT to embed watermarks in perceptually salient yet stable regions within the reverse diffusion process, ensuring robustness without compromising audio quality. By bridging signal structure and the adaptability of neural networks, Smark forms a hybrid framework that achieves superior performance in both audio quality and watermark robustness metrics.

\begin{figure*}[htbp!]
    \centering
    \includegraphics[width=\linewidth]{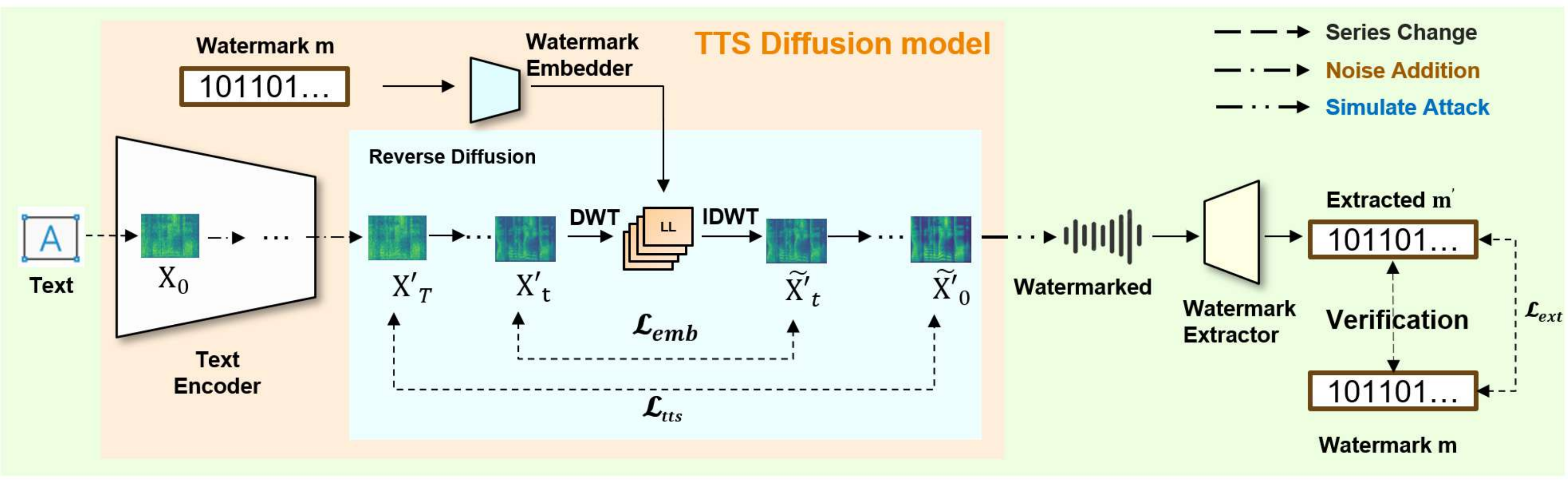}
    \caption{The pipeline of Smark. First, during the reverse diffusion process of a TTS diffusion model, the Mel spectrogram \(X'_t\) at timestep $t$  is decomposed to four sub-bands LL, LH, HL, HH using DWT. Next, the watermark $\mathbf{m}$ is embedded into the LL sub-band of \(X'_t\) via a watermark embedder (see Fig. \ref{fig:Watermark Embedder and Extractor}).  Then, 
    the watermarked LL sub-band is combined with the unmodified LH, HL, HH sub-bands to reconstruct the watermarked Mel spectrogram \(\tilde{X}'_t\) via the inverse transform IDWT, which is passed to the next reverse diffusion step. After the diffusion process is complete, the watermark $\mathbf{m}'$ is extracted using a learned extractor and compared to the original $\mathbf{m}$ for verification, under both the no-attack and various simulated real-world attack scenarios.  }
    \label{fig:pipeline}
\end{figure*}

\section{Preliminaries}
\label{sec:Preliminaries}
\subsection{TTS Diffusion Models}
We briefly review the basics of speech generation process of TTS diffusion models. 
It can be described in two stages. The first stage converts text into an intermediate representation, the Mel spectrogram~\cite{davis1980comparison}, which is a time-frequency representation aligned with human auditory perception. The second stage employs a reverse diffusion process to convert this Mel spectrogram into a generated audio output. These stages correspond to the two core components of all TTS diffusion models: a text encoder responsible for processing text conditions, and a reverse diffusion module responsible for audio generation. While different TTS diffusion models have different structures for the text encoder, the reverse diffusion process shares a common paradigm. This is because they all rely on the fundamental principles of diffusion models.

The training objective of TTS diffusion models typically involves minimizing the variational lower bound on the negative log-likelihood, which can be reduced to a simple mean-squared error between the predicted and actual noise components. This training paradigm ensures stable convergence and high-quality generation. Importantly, the shared mathematical foundation across different TTS diffusion models means that the reverse diffusion process, regardless of specific architectural implementations, follows the same fundamental principles. This commonality provides the theoretical basis for Smark's model-agnostic approach, as the watermark embedding operates within this shared reverse process rather than relying on model-specific components.
The reverse diffusion process is mathematically modeled as a Markov process~\cite{ho2020denoising}, with a conditional probability model of
\begin{equation*}
p_\theta(x_{t-1} | x_t) = \mathcal{N}(x_{t-1}; \mu_\theta(x_t, t), \Sigma_\theta(x_t, t))
\end{equation*}
where $\mu_\theta(x_t, t)$ and $\Sigma_\theta(x_t, t)$ are the predicted mean and covariance of the generative process, respectively. All TTS diffusion models build on this mathematical model to construct the reverse diffusion process, which involves training a neural network to predict \(\mu_\theta(x_t, t)\) and \(\Sigma_\theta(x_t, t)\) for denoising.

\subsection{Discrete Wavelet Transform}

As a time-frequency multi-resolution analysis tool~\cite{davis1980comparison}, Discrete Wavelet Transform (DWT) has shown unique advantages in the processing of non-stationary signals by virtue of its characteristic of ``combining time localization and frequency selectivity". Its core value lies in its ability to decompose signals into low-frequency approximation components (LL sub-band) and high-frequency detail components (LH, HL, HH sub-bands): the low-frequency LL sub-band carries the core structural information of the signal and has strong resistance to interference such as noise and compression; the high-frequency sub-bands mainly contain the detailed features of the signal and are vulnerable to external attacks.

The mathematical formulation of DWT provides insight into its suitability for audio watermarking. 
Given a discrete-time signal \(s_n\) of length \(M\), where \(n\) is the time index, the DWT decomposition can be expressed as
\begin{align*}
&W_{\phi}(j_0,k) = \frac{1}{\sqrt{M}}\sum_n s_n\phi_{j_0,k}[n]\\
&W_{\psi}(j,k) = \frac{1}{\sqrt{M}}\sum_n s_n\psi_{j,k}[n] \quad \text{for } j \geq j_0
\end{align*}
where \(\phi_{j,k}[n]\) and \(\psi_{j,k}[n]\) represent the scaling and wavelet functions at scale \(j\) and position \(k\), respectively. The approximation coefficients \(W_{\phi}\) capture low-frequency components that correspond to the fundamental structure of the signal, while the detail coefficients \(W_{\psi}\) represent higher-frequency information. In the context of Mel spectrograms, this multi-resolution decomposition enables selective watermark embedding in perceptually significant yet stable regions, providing an optimal balance between robustness and imperceptibility.

In speech noise reduction tasks, researchers can retain the core semantic features of speech while removing interference by preserving the LL sub-band~\cite{schroter2022deepfilternet} and suppressing high-frequency noise components. In speech feature extraction tasks, the LL sub-band has been proven to be highly aligned with the sensitive frequency bands of the human auditory system~\cite{sanli2025low}, which can effectively improve the accuracy of tasks such as speech recognition and emotion classification.

We note that the Mel spectrogram in TTS diffusion models can essentially be regarded as a two-dimensional time-frequency image, and its low-frequency LL sub-band also carries the core perceptual features of speech, which has natural advantages for watermark embedding.
Therefore, combining DWT with the reverse diffusion process of TTS diffusion models and embedding watermarks through the LL sub-band is expected to break through the versatility and sound quality limitations of existing methods and provide a new path for the watermarking design of TTS diffusion models.

\section{Proposed Method}
\label{sec:proposed method}

\subsection{At a Glance}
To establish a universal and high-performance watermarking framework for TTS diffusion models, our core idea is to leverage the multi-resolution properties of DWT to embed an imperceptible and robust watermark during the common reverse diffusion process shared by all TTS diffusion models. The proposed Smark framework is shown in Fig.~\ref{fig:pipeline}. 

To achieve it, we use DWT to decompose the Mel spectrogram $X^{'}_t$ at each reverse diffusion step $t$ into four orthogonal sub-bands: the low-frequency approximation (LL) and the high-frequency details (LH, HL, HH). The LL sub-band carries perceptually dominant and structurally stable components, encapsulating the fundamental formants and prosodic contours of speech, whereas high-frequency information is more susceptible to attenuation during both the denoising process and common audio processing operations. 
Therefore, it is intuitive to embed watermarks exclusively in the LL sub-band, which enables seamless integration with the core acoustic structure, improves robustness throughout the iterative diffusion process, and minimizes audible artifacts by avoiding perceptually sensitive high-frequency regions.

\begin{figure}[h]
    \centering
\includegraphics[width=\linewidth]{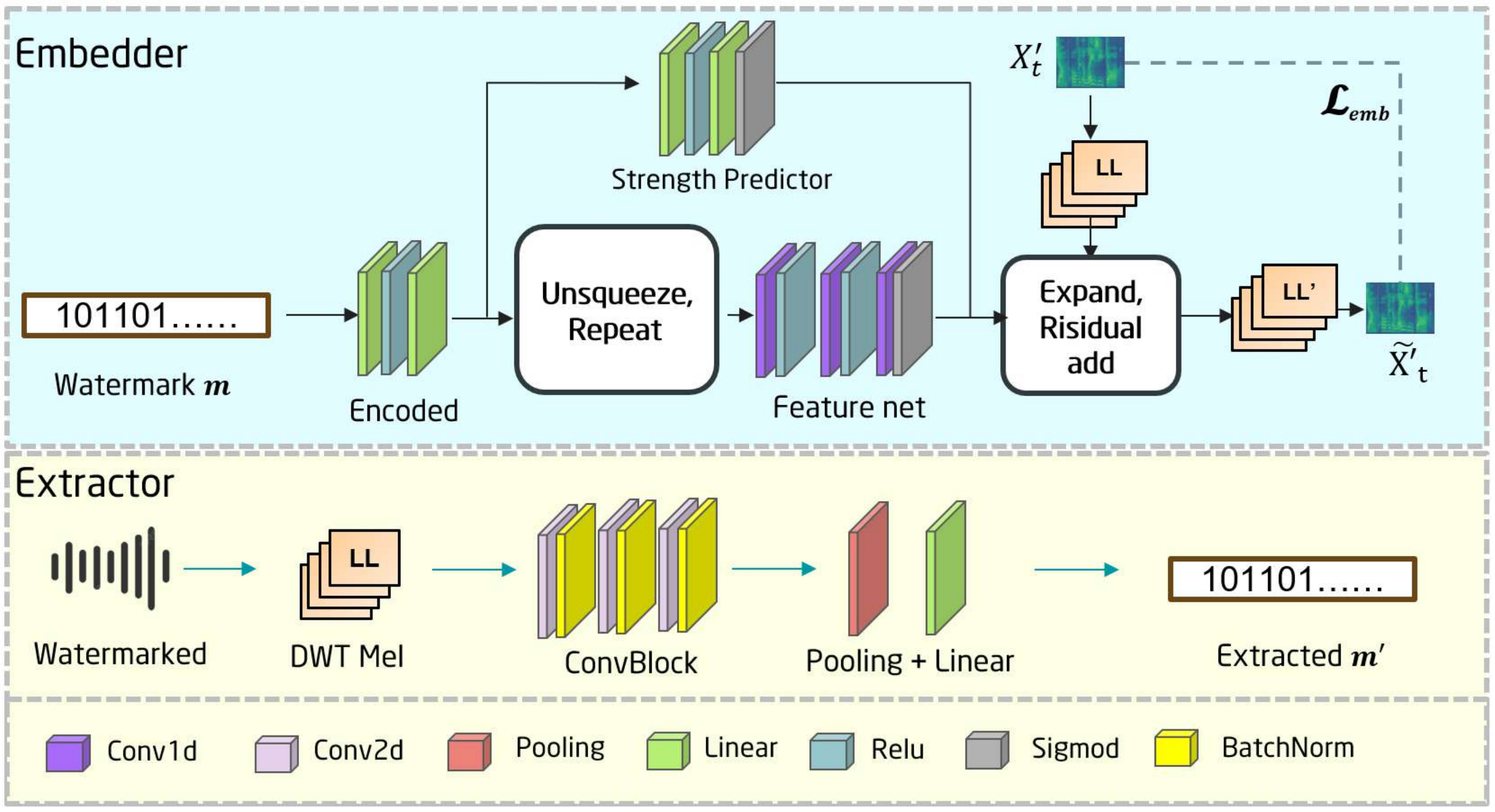}
    \caption{Watermark embedder and extractor}
    \label{fig:Watermark Embedder and Extractor}
\end{figure}

We design a lightweight neural-network based watermark embedder and extractor (see Fig. \ref{fig:Watermark Embedder and Extractor}) to implement the watermarking scheme. A joint optimization strategy is employed to balance audio quality with watermark extraction accuracy, coordinating the training of the TTS model, embedder, and extractor. Furthermore, we simulate various real-world attacks on the generated audio to evaluate the robustness of the embedded watermark, ensuring practical applicability.

\subsection{Watermark Embedder and Extractor}

The watermark embedder fuses a binary watermark sequence 
$\mathbf{m}=(m_1,\ldots,m_N)$ with the intermediate Mel spectrogram of the diffusion model, where $N$ denotes the watermark capacity. After the diffusion process, the watermark is  extracted from the generated audio and compared with the original sequence to verify both its presence and correctness.

As shown in Fig.~\ref{fig:Watermark Embedder and Extractor}, the embedder includes an encoding layer, a feature enhancement network, and an intensity predictor. The input watermark is first encoded into a latent vector via two fully connected layers with ReLU, then expanded temporally through a convolutional feature enhancement network to extract local features and output a single-channel watermark. An intensity predictor generates dynamic coefficients, and the convolved watermark is expanded to match the speech feature dimensions and fused with the original feature residuals.
\begin{align*}
\tilde{X}'_t = X'_t + \alpha \cdot (2\cdot Emb(\mathbf{m}) - 1),\ \ \alpha \in \mathbb{R}
\end{align*}
where $X^{'}_t$ and $\tilde{X}'_t$ are the original and watermarked Mel spectrogram, respectively; 
$\textit{Emb}(\mathbf{m})$ is the watermark representation generated by the watermark embedder, and
$\alpha$ is the dynamic embedding strength coefficient.

The design of the watermark embedder is motivated by the need to balance embedding strength with perceptual transparency. The encoding layer transforms the binary watermark sequence into a latent representation that can be fused with the Mel spectrogram. The feature enhancement network, composed of dilated convolutional layers, captures multi-scale contextual information to ensure that the watermark is distributed across both time and frequency dimensions. The intensity predictor dynamically adjusts the embedding strength 
$\alpha$ based on the local characteristics of the speech signal, thereby allowing stronger embedding in perceptually masked regions and weaker embedding in sensitive regions. This adaptive strategy minimizes audible artifacts while maintaining robust watermark extraction.

The watermark extractor is a simple 2D convolutional neural network consisting of three Convblocks, a pooling layer, and a linear layer. The Convblock consists of a 2D convolutional layer and a normalization function. The Convblock captures local spectral features, the pooling layer compresses the spatial dimensions, and the linear layer maps the convolutional features to the watermark output.

\subsection{Optimization Strategy}
To achieve a balance between audio quality and watermark extraction accuracy,
the Smark framework utilizes a joint optimization strategy of the watermark embedder, the extractor, and the underlying TTS diffusion model.

\textbf{1) Embedder loss.} The embedder aims to optimize its neural network parameters $\theta$ by minimizing the LL sub-band distortion between the original and watermarked Mel spectrograms 
\begin{align*}
\mathcal{L}_{emb}(\theta) &= \frac{1}{H \times W} 
\sum_{h=1}^{H} \sum_{w=1}^{W} \left( {X}^{\text{LL}}_{t}[h,w] - \tilde{X}^{\text{LL}}_{t}[h,w; \theta] \right)^2
\label{eq:loss_emb}
\end{align*}
where \({X}^{\text{LL}}_{t}\)
and 
\(\tilde{X}^{\text{LL}}_{t}\)
are the LL sub-band features 
of the original and watermarked Mel spectrogram at timestep \(t\) of the reverse diffusion process, respectively; \(H \times W\) is the spatial dimension of the LL sub-band; $h$ indexes the frequency bins, and $w$ indexes the time frames.

\textbf{2) Extractor loss.} The watermark extractor aims to optimize its neural network parameters $\zeta$ by minimizing the difference between the extracted watermark and the original watermark
\begin{align*}
\mathcal{L}_{ext}(\zeta) = \text{BCE}(\mathbf{m}, \mathbf{m}'(\zeta))
\end{align*}
where BCE is the binary cross-entropy \cite{de2005tutorial}; \(\mathbf{m}\) is the original binary watermark sequence; \(\mathbf{m}'(\zeta)\) is the extracted sequence.

\textbf{3) Total loss.} The total loss function of Smark method is
\begin{equation*}
\mathcal{L}_{total} =  \lambda_{tts} \mathcal{L}_{tts}
+ \lambda_{emb} \mathcal{L}_{emb} + \lambda_{ext} \mathcal{L}_{ext} 
\end{equation*}
where $\lambda_{tts}$,
$\lambda_{emb}$, $\lambda_{ext}$ are hyper-parameters that balance the contribution of each component to the total training objective.

The joint optimization process employs adaptive weighting of the loss components based on training progress. Initially, higher weight is given to $\mathcal{L}_{tts}$ to ensure stable TTS model convergence. As training progresses, the weights gradually shift toward $\mathcal{L}_{emb}$ and $\mathcal{L}_{ext}$ to refine the watermark embedding and extraction capabilities. This curriculum learning approach prevents interference between the different objectives and ensures coordinated optimization of all components. Additionally, the embedding strength parameter $\alpha$ is dynamically adjusted during training using a scheduled annealing strategy, starting with larger values to establish robust watermark patterns and gradually reducing to minimize perceptual impact.

\subsection{Computational Overhead}
The proposed watermarking framework introduces minimal computational overhead to the TTS diffusion model. Watermark embedding is performed only once at a predefined diffusion timestep and therefore does not scale with the number of diffusion steps. At this timestep, a fixed DWT is applied to an intermediate audio representation, and watermark embedding is conducted in the LL sub-band using a lightweight neural network. As both the DWT and inverse DWT are fixed linear transforms without learnable parameters, their computational cost is linear with respect to the signal length.

The watermark embedder adopts a shallow and parameter-efficient architecture, as illustrated in Fig.~\ref{fig:Watermark Embedder and Extractor}. It operates on reduced-resolution representations, introducing only a small number of additional parameters. After inverse DWT, the diffusion process proceeds normally without altering the remaining diffusion steps, and the inference pipeline remains unchanged.

Watermark extraction is performed offline on the final generated audio using a simple 2D convolutional neural
network extractor and does not affect speech generation efficiency. Overall, the additional computational cost introduced by the watermark embedding and extraction modules is marginal compared to that of the backbone diffusion model.

\subsection{Simulated Attack Scenarios}
\label{sec:robustness}

To simulate real-world scenarios where generated audio might be processed or distorted, we evaluate the watermarking methods against a diverse suite of audio attacks [5, 17]. These attacks are applied to the generated waveform after the TTS synthesis is complete, testing the watermark's survivability, including:

\noindent
\text{\ }1) amplitude clipping (to simulate signal saturation); \\
\noindent
\text{\ }2) addition of 20\%-intensity Gaussian noise (to mimic quality degradation during recording or transmission); \\
\noindent 
\text{\ }3) time-domain masking (randomly zeroing out 10\% of samples to simulate data packet loss); \\
\noindent 
\text{\ }4) 90\% amplitude scaling (to verify sensitivity to volume normalization operations); \\
\noindent 
\text{\ }5) echo effect (introducing attenuated and shifted copies of the Mel spectrogram); \\
\noindent 
\text{\ }6) low-pass filtering with a cutoff frequency of 5000Hz (to weaken potential high-frequency watermark components).

Expect for the above individual attacks,
we also combine them into composite attacks to validate the robustness of Smark under complex, multi-step processing environments. Specifically, we apply two, three even four randomly ordered attacks successively to the same audio sample to simulate the multiple stages of signal processing or quality degradation that audio may undergo during distribution and usage. This setup better reflects the sequential distortion processes in real-world scenarios, thereby comprehensively assessing the survivability and stability of the watermark embedded by Smark under demanding conditions.

\subsection{Watermark Verification}

To statistically validate the effectiveness of watermark embedding and extraction, we adopt a hypothesis testing framework based on the binomial distribution, treating each bit extraction as an independent trial. Let \( N \) denote the watermark length. The number of correctly extracted bits \( X \) follows a binomial distribution with success probability \( \xi \):  
\[
\Pr(X = i) = \binom{N}{i} \xi^i (1-\xi)^{N-i}.
\]  
Then watermark verification is formulated as a one-sided hypothesis test to determine whether the observed number of matching bits significantly exceeds the level expected from random guessing. The hypotheses are defined as:  
\begin{itemize}  
    \item \textit{Null hypothesis (\(H_0\))}: \(\xi = 0.5\), corresponding to random guessing, \textit{i.e.}, no watermark embedded.  
    \item \textit{Alternative hypothesis (\(H_1\))}: \(\xi > 0.5\), indicating successful watermark embedding.  
\end{itemize}

\begin{figure}[t]
    \centering
\includegraphics[width=\linewidth]{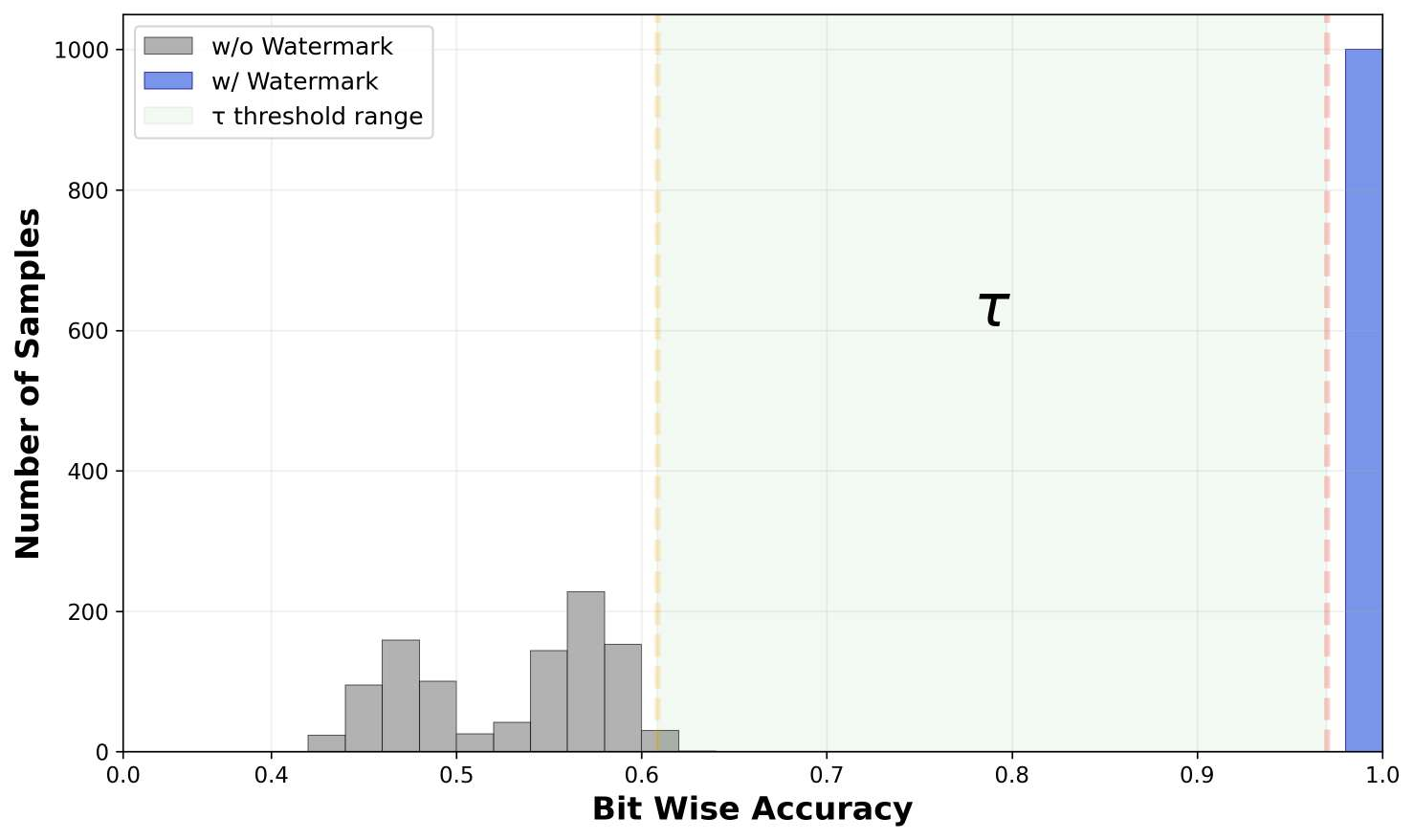}
\caption{A test result for binomial distribution under hypotheses $H_0$ and $H_1$, with watermark length $N=100$ and sample size 1000. For $H_0$, the bit wise accuracy is $\hat{\xi}\approx 0.5331$ (i.e., the average of gray bars); for $H_1$, $\hat{\xi}\approx 0.9983$ (i.e., the average of blue bars). 
The verification threshold $\tau \in [0.62, 0.97]$ ensures both low FPR and low FNR.
} 
\label{fig:Hypothesis testing}
\end{figure}

To make a binary decision on watermark presence, a verification threshold \(\tau \) on bit accuracy is required, where \(\tau\ = k/N\) and $k$ is the critical number of correctly extracted bits. It defines the decision rule: reject \( H_0 \) (and conclude a watermark is present) if and only if the observed bit-match ratio is no less than \( \tau \). Within this framework we consider two types of error.
\begin{itemize}
    \item A \textit{false positive (FP)} is the error of rejecting \( H_0 \) (declaring a watermark) when it is true (no watermark is embedded). This corresponds to placing confidence in a spurious detection.
    \item A \textit{false negative (FN)} is the error of accepting \( H_0 \) (declaring no watermark) when \( H_1 \) is true (a watermark is indeed embedded). This corresponds to missing a genuine watermark.
\end{itemize}

In practice, the threshold $\tau$ is determined by constraining the false positive rate (FPR)
to a sufficiently low level.
Under $H_0$, the number of correctly extracted bits
$X$ follows a binomial distribution $X \sim \mathrm{Binomial}(N, \xi)$.
The FPR corresponding to a given threshold $k$ is computed as
\[
\mathrm{FPR} = \Pr(X \ge k \mid H_0)
= \sum_{i=k}^{N} \binom{N}{i} \xi^i (1-\xi)^{N-i} .
\]
The smallest $k$ that satisfies the desired FPR constraint is selected,
yielding the verification threshold $\tau = k/N$.
Once $\tau$ is fixed, the false negative rate (FNR) under $H_1$
can be evaluated to assess the reliability of watermark detection.

Fig.~\ref{fig:Hypothesis testing} illustrates the binomial distributions under both hypotheses, based on experiments using two sets of 1000 non-watermarked and 1000 watermarked audio samples, respectively.
For $H_0$, the average bit-wise extraction accuracy of non-watermarked samples (gray bars) is $\hat{\xi}\approx
0.5331$,
which is close to random guessing.
For $H_1$, the average bit-wise extraction accuracy of watermarked samples (blue bars) reaches $\hat{\xi}\approx 0.9983>0.5$.
Accordingly, a feasible range of decision thresholds is
approximately $\tau \in [0.62, 0.97]$, within which both FP and FN errors are well controlled. In fact, with $N = 100$, the threshold $\tau = 0.62$ is obtained by enforcing
$\mathrm{FPR}=\Pr(X \ge 62 \mid H_0) \le 0.0017$. On the other hand, 
the empirical results show that the extraction accuracy under $H_1$
is consistently concentrated above $0.97$,
which gives $\mathrm{FNR} = \Pr(X < 97 \mid H_1) \le 0.01$.
These results indicate that the proposed verification rule
achieves highly reliable watermark detection with strong statistical confidence.


\section{Experiments}
\label{sec:experiments}
\subsection{Experiment Setup}

\textbf{1) Datasets.} We employ three mainstream and widely-used datasets to evaluate Smark's capability: 
LJSpeech~\cite{ljspeech17}, LibriTTS~\cite{zen2019libritts}, and LibriSpeech~\cite{panayotov2015librispeech}. 
The LJSpeech dataset~\cite{ljspeech17} provides 13,100 high-quality single-speaker recordings, offering controlled conditions for assessing fundamental performance. Its 22.05 kHz sampling rate and consistent recording environment establish a reliable baseline. In contrast, the LibriTTS dataset~\cite{zen2019libritts} encompasses 585 hours of multi-speaker audio with 24 kHz sampling, introducing real-world complexity through diverse speaking styles, acoustic conditions, and background variations. In addition, the LibriSpeech dataset~\cite{panayotov2015librispeech} shares a similar large-scale, multi-speaker foundation with LibriTTS, but focuses on read speech rather than TTS-generated audio, thus covering a distinct real-world speech.  These three datasets ensure comprehensive evaluation in both ideal and challenging scenarios.

\textbf{2) Baseline methods.} We compare with representative baseline methods covering diverse watermarking paradigms to verify Smark's competitive advantage. WavMark~\cite{chen2023wavmark}, TimbreWM~\cite{liu2023detecting}, DeAR~\cite{li2022dear}, and AudioSeal~\cite{roman2024proactive} exemplify post-processing approaches, operating independently of the audio generation pipeline with mature implementation frameworks. In contrast, GROOT~\cite{liu2024groot} represents state-of-the-art generative watermarking, modifying initial conditions in the diffusion process to achieve intrinsic embedding. This cross-paradigm selection enables systematic comparison across methodological boundaries and validates Smark's universal applicability across different watermarking design logics.

\textbf{3) Evaluation metrics.} We adopt a multi-faceted evaluation strategy capturing both perceptual quality and technical performance.\\ 
\noindent    
\text{\quad}a) Perceptual Evaluation of Speech Quality (PESQ) metric \cite{rix2001perceptual} quantifies speech quality degradation through psychoacoustic modeling.It is used to estimate subjective speech quality scores, providing an automatic approximation of human perceptual judgments. \\
\noindent    
\text{\quad}b) Short-Time Objective Intelligibility (STOI)\cite{taal2010short} specifically assesses intelligibility preservation by analyzing temporal envelope correlations. \\
\noindent    
\text{\quad}c) Mean Opinion Score net (MOSnet) \cite{lo2019mosnet} provides automated subjective quality assessment, bridging objective measurements and human perception. 

For watermark reliability, ACC measures the proportion of exact matches between the embedded and extracted sequences, with statistical significance validated using a binomial hypothesis test at $\alpha=0.01$.

\begin{table}[t]
\centering
\renewcommand{\arraystretch}{1.05}
\caption{Comparison of Smark and baselines across different models and datasets without attack
}
\resizebox{\linewidth}{!}{
\begin{tabular}{llcccc}
\toprule
\textbf{Dataset} & \textbf{Method (Base Model)} & \textbf{PESQ↑} & \textbf{STOI↑} & \textbf{MOS↑} & \textbf{ACC↑} \\
\midrule
\multirow{8}{*}{LJSpeech} 
& AudioSeal~\cite{roman2024proactive} (PriorGrad) & 1.7793 & 0.8931 & 1.5762 & 0.7983 \\
& \textbf{Smark} (PriorGrad) & 2.0856 & 0.9705 & 2.1541 & 0.9863 \\
\cmidrule{2-6}
& GROOT \cite{liu2024groot} (Wavegrad) & 3.4325 & 0.9731 & 1.9635 & 1.0000 \\
& DeAR\cite{li2022dear} (Wavegrad) & 3.1360 & 0.7508 & 1.8336 & 1.0000 \\
& \textbf{Smark} (Wavegrad) & 4.5462 & 0.9957 & 2.0924 & 1.0000 \\
\cmidrule{2-6}
& Wavmark \cite{chen2023wavmark} (GradTTS) & 3.4325 & 0.9692 & 2.9868 & 0.9998 \\
& TimbreWM\cite{liu2023detecting} (GradTTS) & 2.7032 & 0.7357 & 1.9916 & 0.9999 \\
& \textbf{Smark} (GradTTS) & 4.5467 & 1.0000 & 3.1134 & 1.0000 \\
\midrule
\multirow{8}{*}{LibriTTS}
& AudioSeal~\cite{roman2024proactive} (PriorGrad) & 1.6842 & 0.8725 & 1.4936 & 0.7815 \\
& \textbf{Smark} (PriorGrad) & 1.9738 & 0.9512 & 2.0479 & 0.9786 \\
\cmidrule{2-6}
& GROOT \cite{liu2024groot} (Wavegrad) & 3.2867 & 0.9543 & 1.8572 & 0.9992 \\
& DeAR\cite{li2022dear} (Wavegrad) & 2.9845 & 0.7319 & 1.7428 & 0.9989 \\
& \textbf{Smark} (Wavegrad) & 4.3289 & 0.9876 & 1.9863 & 0.9998 \\
\cmidrule{2-6}
& Wavmark \cite{chen2023wavmark} (GradTTS) & 3.2719 & 0.9485 & 2.8735 & 0.9998 \\
& TimbreWM\cite{liu2023detecting} (GradTTS) & 2.5683 & 0.7149 & 1.9961 & 0.9999 \\
& \textbf{Smark} (GradTTS) & 4.2593 & 0.9987 & 3.0276 & 0.9999 \\
\midrule
\multirow{8}{*}{LibriSpeech}
& AudioSeal~\cite{roman2024proactive} (PriorGrad) & 1.6523 & 0.8614 & 1.4725 & 0.7736 \\
& \textbf{Smark} (PriorGrad) & 1.9215 & 0.9034 & 2.0167 & 0.9748 \\
\cmidrule{2-6}
& GROOT \cite{liu2024groot} (Wavegrad) & 3.2416 & 0.9325 & 1.8234 & 0.9917 \\
& DeAR\cite{li2022dear} (Wavegrad) & 2.9317 & 0.7205 & 1.7136 & 0.9908 \\
& \textbf{Smark} (Wavegrad) & 4.2815 & 0.9632 & 1.9547 & 0.9995 \\
\cmidrule{2-6}
& Wavmark \cite{chen2023wavmark} (GradTTS) & 3.2218 & 0.9376 & 2.8415 & 0.9967 \\
& TimbreWM\cite{liu2023detecting} (GradTTS) & 2.5134 & 0.7026 & 1.9614 & 0.9986 \\
& \textbf{Smark} (GradTTS) & 4.1923 & 0.9754 & 2.9815 & 0.9997 \\
\bottomrule
\end{tabular}
}
\label{tab:merged_dataset_results}
\end{table}

\begin{figure*}[htbp!]
    \centering
\includegraphics[scale=0.3]
{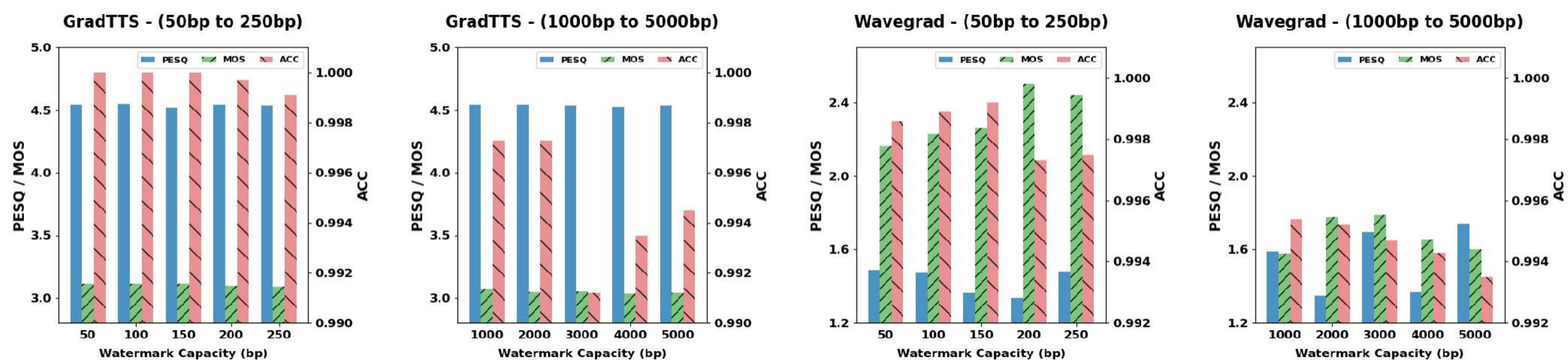}
    \caption{Fidelity of Smark across capacities on LJSpeech for GradTTS and WaveGrad (no attack)}
\label{fig:Capacity}
\end{figure*}

\begin{table*}[h] 
  \centering
  \caption{Comparison of Smark and baselines under individual/hybrid attacks across models on LJSpeech}
   \setlength{\tabcolsep}{6pt}
   \renewcommand{\arraystretch}{0.9}
\setlength{\extrarowheight}{0.5pt}
  \begin{tabular}{crcccccccccc}
    \toprule
    \centering Method (model) & & CLP & Noise-W35 & SS-01 & AS-90 & EA-0315 & LP5000 & CLP+Noise & SS+AS & EA+LP & CLP+AS\\
    \midrule
    \multirow{4}{*}{\shortstack{AudioSeal~\cite{roman2024proactive}\\(Priorgrad)}} & PESQ $\uparrow$& 1.7735 & 1.7521 & 1.7863 & 1.7345 & 1.7689 & 1.7432 & 1.7567 & 1.7701 & 1.7398 & 1.7654 \\& STOI $\uparrow$& 0.8832 & 0.8712 & 0.8654 & 0.8901 & 0.8789 & 0.8856 & 0.8734 & 0.8698 & 0.8821 & 0.8845 \\& MOS $\uparrow$& 1.5604 & 1.5321 & 1.5789 & 1.5234 & 1.5567 & 1.5410 & 1.5378 & 1.5654 & 1.5298 & 1.5576 \\& ACC $\uparrow$& 0.7962 & 0.7854 & 0.7732 & 0.7901 & 0.7812 & 0.7789 & 0.7845 & 0.7723 & 0.7867 & 0.7934\\
        \midrule
        \multirow{4}{*}{\shortstack{\textbf{Smark} \\(Priorgrad)}} & PESQ $\uparrow$ &
    2.0821 & 2.0537 & 2.1245 & 1.9763 & 2.0912 & 2.1148 & 2.0576 & 2.1302 & 1.9935 & 2.0789\\
    & STOI $\uparrow$ & 
    0.9651 & 0.9582 & 0.9471 & 0.9730 & 0.9524 & 0.9692 & 0.9493 & 0.9435 & 0.9601 & 0.9642\\
    & MOS $\uparrow$ & 
    2.1153 & 2.0345 & 2.1923 & 2.0589 & 2.1210 & 2.0756 & 2.1432 & 2.0178 & 2.1045 & 2.0876\\
    & ACC $\uparrow$ & 
    \textbf{0.9801} & \textbf{0.9775} & \textbf{0.9762} & \textbf{0.9830} & \textbf{0.9751} & \textbf{0.9783} & \textbf{0.9805} & \textbf{0.9744} & \textbf{0.9792} & \textbf{0.9821}\\ 
    
\specialrule {0.15em}{1pt}{1pt}
    \multirow{4}{*}{\shortstack{GROOT\cite{liu2024groot}\\(Wavegrad)}} & PESQ $\uparrow$ & 
    3.4472 & 3.4586 & \textbf{3.9574} & 3.5279 & 3.6137 & 3.2586 & 3.4328 & 3.3495 & 3.2478 & 3.5422\\
    & STOI $\uparrow$ & 
    0.9351 & 0.9526 & 0.9827 & 0.8952 & 0.9271 & 0.9665 & 0.9682 & 0.9621 & 0.8948 & 0.9567\\
    & MOS $\uparrow$ & 
    1.9875 & 2.0247 & 2.0195 & 2.9943 & 1.9866 & 2.0422 & 1.9981 & 1.9752 & 1.9882 &1.9786\\
    & ACC $\uparrow$ & 
    0.9921 & 0.9954 & 0.9918 & 0.9917 & 0.9896 & 0.9867 & 0.9817 & 0.9792 &0.9847 & 0.9916\\
        \midrule
    \multirow{4}{*}{\shortstack{DeAR\cite{li2022dear}\\(Wavegrad)}} & PESQ $\uparrow$ & 
    3.0572 & 3.0686 & 3.5674 & 3.1379 & 3.2237 & 2.8686 & 3.0428 & 2.9595 & 2.8578 & 3.1522\\
    & STOI $\uparrow$ & 
    0.6848 & 0.7032 & 0.7403 & 0.6283 & 0.6869 & 0.7229 & 0.7177 & 0.7137 & 0.6584 & 0.7157\\
    & MOS $\uparrow$ & 
    1.6978 & 1.9294 & 1.8313 & 1.6713 & 1.6888 & 1.9244 & 1.7138 & 1.6946 & 1.7047 & 1.6979\\
    & ACC $\uparrow$ & 
    0.9999 & 0.9999 & 1.0000 & 1.0000 & 0.9998 & 0.9993 & 0.9997 & 0.9995 & 0.9994 & 0.9989\\
    \midrule
\multirow{4}{*}{\shortstack{\textbf{Smark} \\(Wavegrad)}} 
    & PESQ $\uparrow$ & \textbf{3.5599} & \textbf{3.8957} & 3.6636 & \textbf{3.8886} & \textbf{3.7843} & \textbf{3.8791} & \textbf{3.6251} & \textbf{3.7706} & \textbf{3.8674} & \textbf{3.7224}\\
    & STOI $\uparrow$ & 0.8831 & 0.9399 & 0.8949 & 0.9376 & 0.9212 & 0.9389 & 0.8921 & 0.9140 & 0.9321 & 0.9038\\
    & MOS $\uparrow$ & 2.3221 & 2.6833 & 2.4293 & 2.6750 & 2.5566 & 2.6680 & 2.3735 & 2.5343 & 2.6492 & 2.4836\\
    & ACC $\uparrow$ & 0.9990 & 0.9880 & 0.9945 & 0.9895 & 0.9918 & 0.9890 & 0.9958 & 0.9926 & 0.9894 & 0.9938\\
\specialrule {0.15em}{1pt}{1pt}
    \multirow{4}{*}{\shortstack{ Wavmark\cite{chen2023wavmark}\\ (GradTTS)}} & PESQ $\uparrow$ & 
    3.6958 & 3.4193 & 3.3125 & 3.6247 & 3.5289 & 3.5826 & 3.7853 & 3.7439 & 3.5173 & 3.5815\\
    & STOI $\uparrow$ & 
    0.9525 & 0.8529 & 0.8426 & 0.9142 & 0.9083 & 0.8727 & 0.8672 & 0.8722 & 0.8796 & 0.9248\\
    & MOS $\uparrow$ & 
    2.8547 & 2.5462 & 2.4581 & 2.7296 & 2.7544 & 2.7492 & 2.6197 & 2.9328 & 2.7361 & 2.7619\\
    & ACC $\uparrow$ & 
    0.9754 & 0.5267 & 0.5364 & 0.8924 & 0.8937 & 0.8259 & 0.8566 & 0.8417 & 0.8419 & 0.9737\\
    \midrule
    \multirow{4}{*}{\shortstack{TimbreWM\cite{liu2023detecting}\\(GradTTS)}} & PESQ $\uparrow$ & 
    2.6892 & 2.6639 & 2.6728 & 2.7249 & 2.3926 & 2.6257 & 2.4824 & 2.5296 & 2.5283 & 2.5295\\
    & STOI $\uparrow$ & 
    0.7090 & 0.7048 & 0.7196 & 0.7123 & 0.6982 & 0.7039 & 0.6960 & 0.6945 & 0.6867 & 0.7029\\
    & MOS $\uparrow$ & 
    1.9521 & 1.9278 & 1.9892 & 1.9159 & 1.6587 & 1.9053 & 1.7095 & 1.8247 & 1.7935 & 1.5218\\
    & ACC $\uparrow$ & 
    0.9992 & 0.9971 & 0.9998 & 0.9918 & \textbf{1.0000} & 0.9993 & 0.9998 & 0.9993 & 0.9997 & 0.9999\\
    \midrule
    \multirow{4}{*}{\shortstack{\textbf{Smark} \\(GradTTS)}} & PESQ $\uparrow$ & 
    4.5475 & 4.5444 & 4.5319 & 4.5475 & 4.5475 & 4.5472 & 4.5474 & 4.5486 & 4.5466 & 4.5472\\
    & STOI $\uparrow$ & 
    1.0000 & 0.9999 & 0.9999 & 0.9998 & 0.9998 & 0.9998 & 0.9998 & 0.9999 & 0.9997 & 0.9999\\
    & MOS $\uparrow$ & 
    3.0550 & 3.0551 & 3.0430 & 3.0555 & 3.0545 & 3.0545 & 3.0550 & 3.0543 & 3.0543 & 3.0548\\
    & ACC $\uparrow$ & 
    \textbf{1.0000} & \textbf{1.0000} & \textbf{1.0000} & \textbf{1.0000} &
    \textbf{1.0000} & \textbf{1.0000} & \textbf{1.0000} &
    \textbf{1.0000} &
    \textbf{1.0000} & \textbf{1.0000}\\

    \bottomrule
  \end{tabular}
  \label{tab:simulation_attack}
\end{table*}

\textbf{4) Implementation details.} 
We conduct experiments on three architecturally distinct TTS diffusion models: GradTTS \cite{popov2021grad}, WaveGrad \cite{chen2020wavegrad} and PriorGrad~\cite{lee2021priorgrad}. GradTTS employs a HiFi-GAN vocoder \cite{kong2020hifi} for waveform synthesis, while WaveGrad utilizes gradient-based sampling for direct waveform generation. PriorGrad incorporates a prior distribution adapted to the input acoustic conditions to accelerate convergence and improve sample quality, providing an additional testbed for evaluating Smark's model-agnostic properties. Training employs the Adam optimizer \cite{zhang2018improved} with learning rate $1\times10^{-4}$ and balanced loss coefficients $\lambda_{tts}=1$, $\lambda_{emb}=2$, $\lambda_{ext}=10$. 
All experiments execute on standardized hardware (Intel i7-11700K, NVIDIA RTX 4090) to ensure reproducible performance measurements.

\subsection{Experiment Results on Fidelity}

Fidelity measures the extent to which watermark embedding affects the perceptual quality of generated speech~\cite{oord2018parallel}. Higher values of PESQ, STOI, and MOS indicate better preservation of speech quality and intelligibility.

Table~\ref{tab:merged_dataset_results} shows the audio fidelity performance of Smark and baseline methods on different models with a fixed watermark capacity of 100bp. Overall, Smark consistently achieves higher objective metrics PESQ and STOI scores than existing methods, indicating its strong ability to preserve audio quality while embedding watermarks. This advantage is attributed to Smark’s strategy of embedding watermarks in perceptually stable regions, which effectively limits interference with critical speech components. 

In addition to objective metrics, Smark attains higher MOS scores than competing approaches. The agreement between subjective and objective evaluations further confirms the reliability of Smark in maintaining audio fidelity.

Meanwhile, evaluations conducted on the single-speaker LJSpeech~\cite{ljspeech17},
multi-speaker LibriTTS~\cite{zen2019libritts}, and the comparative LibriSpeech dataset~\cite{panayotov2015librispeech} show that, despite differences in speaker diversity, recording conditions, and speech characteristics, Smark is able to maintain stable performance with minimal degradation.

Furthermore, Table~\ref{tab:simulation_attack} reports the comparison of audio
fidelity under various simulated attack conditions. Even under simulated attacks targeting audio signals, Smark generally achieves higher metrics than baseline methods and shows its ability to preserve high audio fidelity in adversarial settings.

\begin{figure*}[t]
    \centering
\includegraphics
[width=\linewidth]
{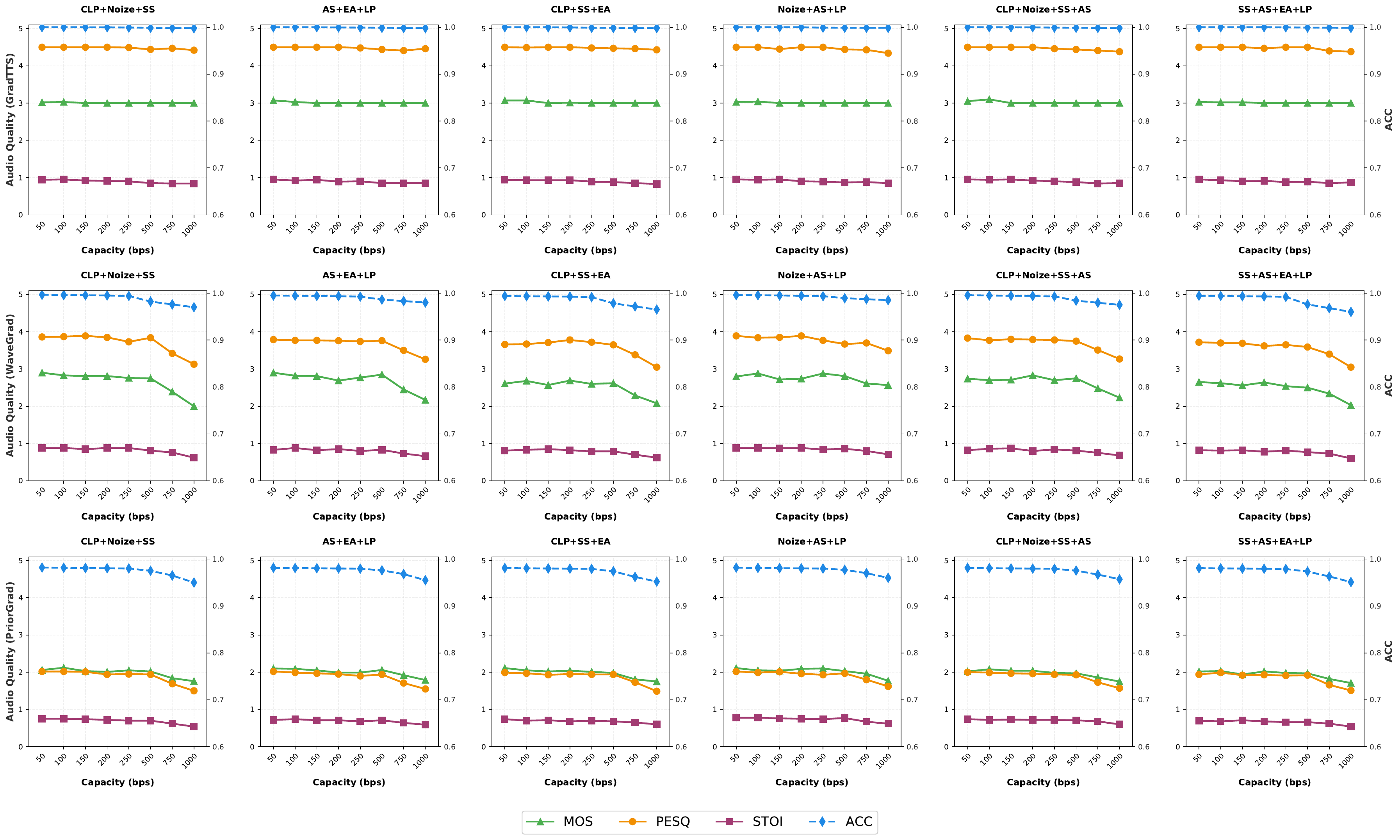}
\caption{Comparison of Smark's performance under composite attacks across different capacities, on GradTTS (first row), WaveGrad (second row), and PriorGrad (third row) using the LJSpeech dataset.}
\label{fig:Different Capacity}
\end{figure*}

\subsection{Experiment Results on Robustness}
\subsubsection{\textit{Robustness on different capacities}} 
Capacity refers to the number of watermark bits that can be appropriately embedded. We evaluate the performance of Smark across a wide range of watermark capacity varying from 50bp to 5000bp in the scenario without attacks.

As shown in Fig.~\ref{fig:Capacity}, Smark shows remarkable stability in audio quality even as watermark capacity scales dramatically from 50bp to 5000bp, while maintaining near-perfect ACC throughout the entire capacity spectrum. Specifically, its PESQ scores and MOS values stay stable without significant fluctuations, showing no obvious downward trend with the increase of watermark payload. Meanwhile, ACC remains reliably high at no less than 0.990 at all tested capacity levels.

These results show Smark's strong ability to preserve audio quality under varying watermarking demands. Even with a 100-fold increase in capacity, audio quality degradation remains minimal, enabling Smark to flexibly support both low and high capacity data-carrying requirements without compromising performance.

\subsubsection
{\textit{Robustness under attacks}} 
As shown in Table~\ref{tab:simulation_attack}, 
Smark consistently achieves the high ACC 
in all attack scenarios. Most audio watermarking schemes face a universal robustness-quality trade-off dilemma: to guarantee high extraction accuracy against signal perturbations, these schemes usually introduce non-negligible degradation of perceptual audio quality; conversely, prioritizing the preservation of audio naturalness will lead to the reduction of watermark robustness. In contrast, generative methods like Smark, leveraging DWT-based low-frequency embedding, achieve PESQ $\ge 3.5$, STOI $\ge 0.88$, and high ACC, balancing robustness and audio quality effectively.

In addition to superior robustness, Smark achieves higher PESQ and STOI scores than existing baselines in most attack scenarios, which shows its superior ability to preserve speech quality while withstanding signal perturbations. These metrics confirm the superior robustness of Smark even under highly challenging conditions, validating that its integration of DWT low-frequency embedding and reverse diffusion process ensures deep fusion between watermark and speech core features, enabling resistance to both single and hybrid attacks while retaining high audio quality.

\subsubsection{\textit{Joint Robustness on different capacities and stronger attacks}}

To further validate the robustness of Smark in more challenging scenarios, we conduct multiple simulated composite attack experiments on different TTS diffusion models, with watermark capacities ranging from 50bp to 1000bp.

As illustrated  in Fig.~\ref{fig:Different Capacity}, Smark shows an exceptionally stable performance on GradTTS. Even at a low watermark capacity of 50bp, Smark consistently achieves near-perfect ACC under all tested composite attack scenarios. 
Moreover, all four evaluation metrics on GradTTS show minimal variation with respect to both watermark capacity and attack intensity, remaining nearly constant across the entire experimental range. Also, as shown in Table~\ref{tab:simulation_attack}, Smark maintains \(\mathrm{ACC}=1.0\) on GradTTS under certain one-level and two-level attacks. Such a stable and saturated performance may be attributed to the strong generative capability and denoising stability of GradTTS, which can mask subtle degradations introduced by watermark embedding and attacks. 

In contrast, on WaveGrad and PriorGrad models, Smark shows a more pronounced yet reasonable performance variation under the joint effects of higher watermark capacity and stronger composite attacks. Specifically, the extraction accuracy decreases from nearly 1.0 to approximately 0.97 and 0.95 on WaveGrad and PriorGrad, respectively, also accompanied by a moderate degradation in audio quality. Notably, these trends are consistent with conventional expectations for audio watermarking systems operating under high-capacity embedding and severe attack conditions, while the extraction accuracy remains within a high-confidence interval. Overall, the experimental results show that Smark achieves strong joint robustness and effectiveness across different watermark capacities, attack intensities, and TTS diffusion models.

\section{Ablation Study}
\label{sec:ablation_study}

\subsection{With and Without DWT}

To validate the necessity of using DWT in the proposed watermarking framework, we conduct  experiments to compare
the performance of two scenarios: ``with DWT” and ``without
DWT”. As illustrated in
Tabel \ref{tab:Abation:w.wo.DWT}, all evaluation metrics of ``w/ DWT” schemes are
significantly superior to those of the corresponding ``w/o DWT” schemes. This observation confirms the
indispensable role of DWT in balancing audio quality and
watermark robustness. 

In fact, this performance gap stems from two aspects. 
First, since DWT’s LL sub-band retains core perceptual information with reduced resolution, the watermark embedding of Smark is confined to the controllable DWT sub-bands to prevent the watermark from appearing in auditory-sensitive regions, thereby reducing the impact on audio quality. Next, the LL sub-band can resist signal processing and diffusion noise better than full-resolution features, ensuring watermark integrity and accuracy. In contrast, embedding in original features (without DWT) degrades audio quality and watermark robustness due
to perceptual distortion and diffusion randomness.

\begin{table}[htbp!]
\centering
\setlength{\tabcolsep}{4pt}
\renewcommand{\arraystretch}{0.95}
\caption{Comparison of Smark's performance with and without DWT.}
\label{tab:Abation:w.wo.DWT}
\begin{tabular}{ccccccc}
\toprule
Dataset & Model & Method & PESQ↑ & STOI↑ & MOS↑ & ACC↑ \\
\midrule
\multirow{6}{*}{LJSpeech} 
& \multirow{2}{*}{GradTTS} 
    & w/ DWT & \textbf{4.5467} & \textbf{1.0000} & \textbf{3.1134} & \textbf{1.0000} \\
&   & w/o DWT & 3.0458 & 0.7653 & 1.6821 & 0.5436 \\
\cmidrule(lr){2-7}
& \multirow{2}{*}{Wavegrad} 
    & w/ DWT & \textbf{4.5462} & \textbf{0.9957} & \textbf{2.0924} & \textbf{1.0000} \\
&   & w/o DWT & 3.2372 & 0.6806 & 1.5612 & 0.6735 \\
\cmidrule(lr){2-7}
& \multirow{2}{*}{Priorgrad} 
    & w/ DWT & \textbf{2.0856} & \textbf{0.9705} & \textbf{2.1541} & \textbf{0.9863} \\
&   & w/o DWT & 1.5057 & 0.5926 & 1.6825 & 0.5082 \\
\midrule
\multirow{6}{*}{LibriTTS}
& \multirow{2}{*}{GradTTS} 
    & w/ DWT & \textbf{4.2593} & \textbf{0.9987} & \textbf{3.0276} & \textbf{0.9999} \\
&   & w/o DWT & 3.3279 & 0.7710 & 1.7352 & 0.5584 \\
\cmidrule(lr){2-7}
& \multirow{2}{*}{Wavegrad} 
    & w/ DWT & \textbf{4.3289} & \textbf{0.9876} & \textbf{1.9863} & \textbf{0.9998} \\
&   & w/o DWT & 3.1268 & 0.6468 & 1.7352 & 0.6308 \\
\cmidrule(lr){2-7}
& \multirow{2}{*}{Priorgrad} 
    & w/ DWT & \textbf{1.9738} & \textbf{0.9512} & \textbf{2.0479} & \textbf{0.9786} \\
&   & w/o DWT & 1.5372 & 0.5613 & 1.5627 & 0.5236 \\
\midrule
\multirow{6}{*}{LibriSpeech}
& \multirow{2}{*}{GradTTS} 
    & w/ DWT & \textbf{4.1923} & \textbf{0.9754} & \textbf{2.9815} & \textbf{0.9997} \\
&   & w/o DWT & 3.3630 & 0.7178 & 1.9253 & 0.6352 \\
\cmidrule(lr){2-7}
& \multirow{2}{*}{Wavegrad} 
    & w/ DWT & \textbf{4.2815} & \textbf{0.9632} & \textbf{1.95
47} & \textbf{0.9995} \\
&   & w/o DWT & 3.1687 & 0.6623 & 1.8268 & 0.5914 \\
\cmidrule(lr){2-7}
& \multirow{2}{*}{Priorgrad} 
    & w/ DWT & \textbf{1.9215} & \textbf{0.9034} & \textbf{2.0167} & \textbf{0.9748} \\
&   & w/o DWT & 1.5194 & 0.5260 & 1.6434 & 0.5126 \\
\bottomrule
\end{tabular}
\end{table}

\begin{figure*}[htbp!]
    \centering
\includegraphics[width=\linewidth]
{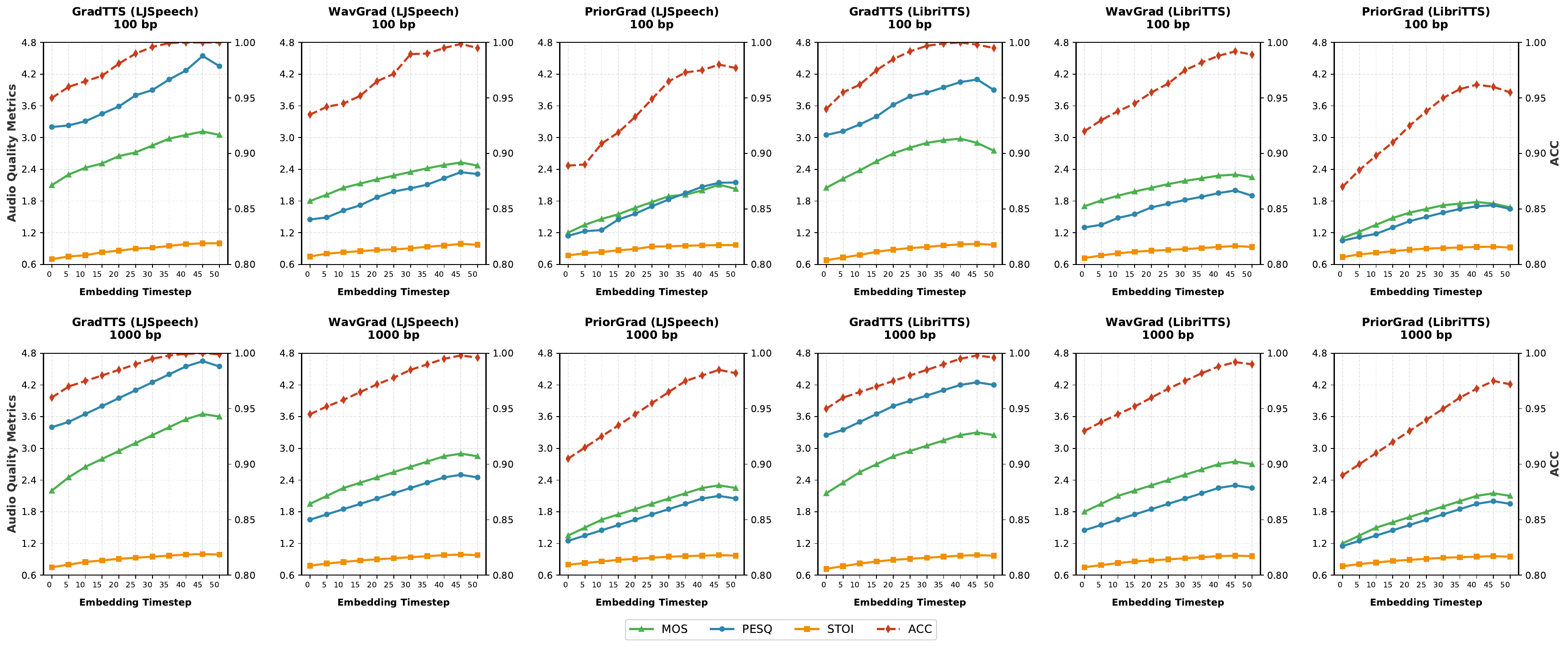}
    \caption{Ablation study of Smark embedding at different timesteps.
    It compares the performance of Smark with different embedding start timesteps (x-axis), ranging from timestep 0 to timestep 50, under two capacity settings: 100bp and 1000bp.} 

\label{fig:Timesteps}
\end{figure*}

\subsection{Embedding in Different DWT Sub-bands}
To validate the effectiveness of embedding watermarks in the low-frequency LL sub-band, we conduct an ablation study by comparing the performance of Smark when watermarks are embedded in different DWT sub-bands.

\begin{table}[htbp!]
\centering
\renewcommand{\arraystretch}{0.96}
\caption{Comparison of Smark's watermark embedding on different DWT sub-bands.}
\label{tab:ablation}
\begin{tabular}{c c c c c c}
\toprule
Model (dataset) & Sub-band & PESQ$\uparrow$ & STOI$\uparrow$ & MOS$\uparrow$ & ACC$\uparrow$ \\
\midrule
\multirow{4}{*}{\makecell{GradTTS\\(LJSpeech)}}
 & LL & \textbf{4.5467} & \textbf{1.0000} & \textbf{3.1134} & \textbf{1.0000} \\
 & LH & 3.0208 & 0.4713 & 1.7212 & 0.5219 \\
 & HL & 3.2643 & 0.4579 & 1.6854 & 0.4496 \\
 & HH & 4.3195 & 0.9850 & 2.1158 & 0.8545 \\
 \cmidrule(lr){2-6}
\multirow{4}{*}{\makecell{GradTTS\\(LibriTTS)}}
 & LL & \textbf{4.2593} & \textbf{0.9987} & \textbf{3.0276} & \textbf{0.9999} \\
 & LH & 2.8376 & 0.4529 & 1.6847 & 0.5138 \\
 & HL & 3.1721 & 0.4384 & 1.6489 & 0.4375 \\
 & HH & 4.0832 & 0.9682 & 2.0643 & 0.8317 \\
 \cmidrule(lr){2-6}
\multirow{4}{*}{\makecell{GradTTS\\(LibriSpeech)}}
 & LL & \textbf{4.1923} & \textbf{0.9754} & \textbf{2.9815} & \textbf{0.9997} \\
 & LH & 2.7914 & 0.4463 & 1.6628 & 0.4926 \\
 & HL & 3.1057 & 0.4301 & 1.6254 & 0.4219 \\
 & HH & 4.0178 & 0.9618 & 2.0336 & 0.8229 \\
\midrule
\multirow{4}{*}{\makecell{WaveGrad\\(LJSpeech)}}
 & LL & \textbf{4.5462} & \textbf{0.9957} & 2.0924 & \textbf{1.0000} \\
 & LH & 3.4257 & 0.6824 & 2.4175 & 0.7246 \\
 & HL & 3.3164 & 0.6258 & 2.2893 & 0.6715 \\
 & HH & 4.2753 & 0.9689 & \textbf{3.0247} & 0.9218 \\
 \cmidrule(lr){2-6}
\multirow{4}{*}{\makecell{WaveGrad\\(LibriTTS)}}
 & LL & \textbf{4.3289} & \textbf{0.9876} & 1.9863 & \textbf{0.9998} \\
 & LH & 3.2145 & 0.6537 & 2.3264 & 0.6982 \\
 & HL & 3.0872 & 0.5946 & 2.1957 & 0.6439 \\
 & HH & 4.0127 & 0.9514 & \textbf{2.9462} & 0.8954 \\
  \cmidrule(lr){2-6}
\multirow{4}{*}{\makecell{WaveGrad\\(LibriSpeech)}}
 & LL & \textbf{4.2815} & \textbf{0.9632} & 1.9547 & \textbf{0.9995} \\
 & LH & 3.1528 & 0.6372 & 2.2845 & 0.6753 \\
 & HL & 3.0146 & 0.5783 & 2.1479 & 0.6217 \\
 & HH & 4.0213 & 0.9457 & \textbf{2.8836} & 0.8872 \\
\midrule
\multirow{4}{*}{\makecell{PriorGrad\\(LJSpeech)}}
 & LL & \textbf{2.0856} & \textbf{0.9705} & \textbf{2.1541} & \textbf{0.9863} \\
 & LH & 1.8724 & 0.5817 & 1.8935 & 0.6728 \\
 & HL & 1.7946 & 0.5429 & 1.8157 & 0.6245 \\
 & HH & 1.9938 & 0.8864 & 2.0472 & 0.9216 \\
  \cmidrule(lr){2-6}
\multirow{4}{*}{\makecell{PriorGrad\\(LibriTTS)}}
 & LL & \textbf{1.9738} & \textbf{0.9512} & \textbf{2.0479} & \textbf{1.0000} \\
 & LH & 1.7853 & 0.5548 & 1.8264 & 0.6519 \\
 & HL & 1.7127 & 0.5186 & 1.7583 & 0.6017 \\
 & HH & 1.8865 & 0.8627 & 1.9648 & 0.9034 \\
  \cmidrule(lr){2-6}
\multirow{4}{*}{\makecell{PriorGrad\\(LibriSpeech)}}
 & LL & \textbf{1.9215} & \textbf{0.9034} & \textbf{2.0167} & \textbf{0.9748} \\
 & LH & 1.7286 & 0.5319 & 1.7842 & 0.6308 \\
 & HL & 1.6574 & 0.4975 & 1.7126 & 0.5879 \\
 & HH & 1.8359 & 0.8378 & 1.9254 & 0.8867 \\
\bottomrule
\end{tabular}
\end{table}

Table~\ref{tab:ablation} presents the audio quality and watermark extraction accuracy when watermarks are embedded in the LL, LH, HL, and HH sub-bands, respectively.

The results show that embedding watermarks in the LL sub-band yields the
best overall performance, achieving near-perfect extraction accuracy while
maintaining superior perceptual quality across all objective metrics.
This behavior can be primarily attributed to the structural stability of the
LL sub-band and the explicit constraint imposed on the watermark embedding strength.
By limiting the embedding magnitude in the approximation sub-band, the introduced
distortions remain subtle and do not cause noticeable degradation in the generated
audio.

Although low-frequency components are usually considered perceptually important,
psychoacoustic studies based on the Discrete Wavelet Packet Transform (DWPT) suggest that small-amplitude modifications in energetically dominant and temporally stable
components can still be perceptually tolerated under signal-dependent conditions
\cite{he2006enhanced}.
Here in the diffusion-based generation process, such constrained low-frequency
perturbations are further regularized in iterative denoising, which helps suppress perceptually salient artifacts.

In contrast, watermark embedding in higher-frequency sub-bands (LH and HL) leads to significant degradation in extraction accuracy and perceptual quality, as these
components are more sensitive to denoising and prone to being smoothed out during the reverse diffusion process. Although the HH sub-band preserves relatively acceptable audio quality, its
extraction accuracy remains substantially lower than that achieved in the LL
sub-band, indicating reduced robustness. Overall, these results confirm that LL sub-band embedding with controlled embedding strength provides a more reliable and perceptually stable watermarking strategy for
TTS diffusion model.

In addition, these results confirm our design intuition: the LL sub-band carries the core structural and perceptual information of speech, providing an ideal balance between imperceptibility and robustness for watermark embedding. The significant performance gap between LL and other sub-bands further justifies our choice of focusing on the low-frequency region for watermark integration.

\subsection{Embedding at Different Timesteps}
To determine the optimal timestep for watermark embedding in the reverse diffusion process, we conduct ablation experiments on GradTTS, WaveGrad, and PriorGrad models with different embedding start points, ranging from timestep 0 to timestep 50, on the LJSpeech and LibriTTS datasets.
Performance is evaluated in terms of both audio quality and watermark ACC, using capacities of 100bp and 1000bp.

As illustrated in Fig.~\ref{fig:Timesteps}, across all model-dataset-capacity combinations, the audio quality metrics consistently show a gradual increase from timestep 0 to timestep around 40-45, reaching a peak, followed by a slight decrease at timestep 50. This pattern indicates that embedding at \textit{intermediate-to-later} timesteps better matches the stable acoustic structure of the Mel spectrogram, thus minimizing perceptual distortion. For the ACC, its values steadily increase from the initial timestep and reach high-level around timestep 35-40 across most configurations. In addition, the capacity 1000bp generally achieves slightly higher accuracy values compared to 100bp.

In other words, despite differences in model architecture, dataset characteristics, and watermark capacity, Smark's watermark embedding in the reverse diffusion process has consistent performance, with intermediate-to-later timesteps providing a more stable and semantically aligned acoustic representation. This consistency further confirms Smark's model-agnostic nature. In all experiments reported in this paper, a timestep of 45 is chosen, as it consistently ensures high audio quality while achieving high-level ACC in all evaluated configurations. It also implies that Smark's design, with watermark embedding in the later stages of the reverse diffusion process, is both computationally efficient and highly effective.

\section{Conclusion}
\label{sec:conclusion}

We proposed \textit{Smark}, a universal watermarking framework for TTS diffusion models. By embedding watermarks into the LL sub-band of Mel spectrograms via DWT during reverse diffusion process, Smark achieves imperceptibility, robustness, and model-agnostic deployment. Experiments on multiple TTS diffusion models show that Smark preserves audio quality and achieves stable and reliable watermark extraction performance, even under various simulated attack scenarios.
Comprehensive ablation studies further analyze the effects of embedding sub-bands, timesteps, and the use of DWT, confirming the importance of low-frequency embedding in maintaining both imperceptibility and robustness. In addition, hypothesis testing is employed to statistically verify the reliability of the watermark detection results. Overall, these results indicate that Smark is an effective and practical solution for speech authentication and intellectual property protection. Future work will explore adaptive watermarking strategies that dynamically adjust to acoustic content, and extend the framework to other generative audio tasks such as music synthesis and audio style transfer.

\bibliographystyle{IEEEtran}
\bibliography{ref}

@inproceedings{liu2024groot,
  title={Groot: Generating robust watermark for diffusion-model-based audio synthesis},
  author={Liu, Weizhi and Li, Yue and Lin, Dongdong and Tian, Hui and Li, Haizhou},
  booktitle={Proceedings of the 32nd ACM International Conference on Multimedia},
  pages={3294--3302},
  year={2024}
}

@inproceedings{zhou2024traceablespeech,
  title={TraceableSpeech: Towards Proactively Traceable Text-to-Speech with Watermarking},
  author={Zhou, Junzuo and Yi, Jiangyan and Wang, Tao and Tao, Jianhua and Bai, Ye and Zhang, Chu Yuan and Ren, Yong and Wen, Zhengqi},
  booktitle={Proceedings of the Interspeech},
  pages={2250--2254},
  year={2024}
}

@inproceedings{popov2021grad,
  title={Grad-tts: A diffusion probabilistic model for text-to-speech},
  author={Popov, Vadim and Vovk, Ivan and Gogoryan, Vladimir and Sadekova, Tasnima and Kudinov, Mikhail},
  booktitle={Proceedings of the International Conference on Machine Learning},
  pages={8599--8608},
  year={2021},
  organization={PMLR}
}

@article{chen2020wavegrad,
  title={Wavegrad: Estimating gradients for waveform generation},
  author={Chen, Nanxin and Zhang, Yu and Zen, Heiga and Weiss, Ron J and Norouzi, Mohammad and Chan, William},
  journal={arXiv preprint arXiv:2009.00713},
  year={2020}
}

@article{wang2023neural,
 author={Chen, Sanyuan and Wang, Chengyi and Wu, Yu and Zhang, Ziqiang and Zhou, Long and Liu, Shujie and Chen, Zhuo and Liu, Yanqing and Wang, Huaming and Li, Jinyu and He, Lei and Zhao, Sheng and Wei, Furu},
  journal={IEEE Transactions on Audio, Speech and Language Processing}, 
  title={Neural Codec Language Models are Zero-Shot Text to Speech Synthesizers}, 
  year={2025},
  volume={33},
  number={},
  pages={705-718},
}

@article{ho2020denoising,
  title={Denoising diffusion probabilistic models},
  author={Ho, Jonathan and Jain, Ajay and Abbeel, Pieter},
  journal={Advances in Neural Information Processing Systems},
  volume={33},
  pages={6840--6851},
  year={2020}
}

@article{davis1980comparison,
  author={Davis, Steven and Mermelstein, Paul},
title={Comparison of parametric representations for monosyllabic word recognition in continuously spoken sentences},  
journal={IEEE Transactions on Acoustics, Speech, and Signal Processing},
  volume={28},
  number={4},
  pages={357--366},
  year={1980},
  publisher={IEEE}
}

@article{kong2020hifi,
  title={Hifi-gan: Generative adversarial networks for efficient and high fidelity speech synthesis},
  author={Kong, Jungil and Kim, Jaehyeon and Bae, Jaekyoung},
  journal={Advances in Neural Information Processing Systems},
  volume={33},
  pages={17022--17033},
  year={2020}
}

@inproceedings{zhang2018improved,
  title={Improved adam optimizer for deep neural networks},
  author={Zhang, Zijun},
  booktitle={Proceedings of the IEEE/ACM 26th International Symposium on Quality of Service (IWQoS)},
  pages={1--2},
  year={2018},
  organization={IEEE}
}

@inproceedings{boney1996digital,
  title={Digital watermarks for audio signals},
  author={Boney, Laurence and Tewfik, Ahmed H and Hamdy, Khaled N},
  booktitle={Proceedings of the third IEEE International Conference on Multimedia Computing and Systems},
  pages={473--480},
  year={1996},
  organization={IEEE}
}

@misc{ljspeech17,
  author       = {Keith Ito and Linda Johnson},
  title        = {The \text{LJSpeech} Dataset},
  howpublished = {\url{https://keithito.com/LJ-Speech-Dataset/}},
  year         = 2017
}

@article{chen2023wavmark,
  title={Wavmark: Watermarking for audio generation},
  author={Chen, Guangyu and Wu, Yu and Liu, Shujie and Liu, Tao and Du, Xiaoyong and Wei, Furu},
  journal={arXiv preprint arXiv:2308.12770},
  year={2023}
}

@article{lo2019mosnet,
  title={MOSNet: Deep Learning-Based Objective Assessment for Voice Conversion},
  author={Lo, Chen-Chou and Fu, Szu-Wei and Huang, Wen-Chin and Wang, Xin and Yamagishi, Junichi and Tsao, Yu and Wang, Hsin-Min},
  journal={Proceedings of the Interspeech},
  year={2019},
  publisher={ISCA}
}

@inproceedings{liu2023detecting,
  title={Detecting voice cloning attacks via timbre watermarking},
  author={Liu, Chang and Zhang, Jie and Zhang, Tianwei and Yang, Xi and Zhang, Weiming and Yu, Nenghai},
  booktitle={Network and Distributed System Security Symposium (NDSS)},
  year={2024}
}

@article{zen2019libritts,
   title     = {LibriTTS: A Corpus Derived from LibriSpeech for Text-to-Speech},
  author    = {Heiga Zen and Viet Dang and Rob Clark and Yu Zhang and Ron J. Weiss and Ye Jia and Zhifeng Chen and Yonghui Wu},
  year      = {2019},
  journal = {Proceedings of the Interspeech},
  pages     = {1526--1530},
  issn      = {2958-1796},
}

@inproceedings{rix2001perceptual,
  title={Perceptual evaluation of speech quality (PESQ)-a new method for speech quality assessment of telephone networks and codecs},
  author={Rix, Antony W and Beerends, John G and Hollier, Michael P and Hekstra, Andries P},
  booktitle={2001 IEEE International Conference on Acoustics, Speech, and Signal Processing},
  volume={2},
  pages={749--752},
  year={2001},
  organization={IEEE}
}

@inproceedings{taal2010short,
  title={A short-time objective intelligibility measure for time-frequency weighted noisy speech},
  author={Taal, Cees H and Hendriks, Richard C and Heusdens, Richard and Jensen, Jesper},
  booktitle={IEEE International Conference on Acoustics, Speech and Signal Processing},
  pages={4214--4217},
  year={2010},
  organization={IEEE}
}

@inproceedings{huang2022prodiff,
  title={Prodiff: Progressive fast diffusion model for high-quality text-to-speech},
  author={Huang, Rongjie and Zhao, Zhou and Liu, Huadai and Liu, Jinglin and Cui, Chenye and Ren, Yi},
  booktitle={Proceedings of the 30th ACM International Conference on Multimedia},
  pages={2595--2605},
  year={2022}
}

@inproceedings{chen2023lightgrad,
  title={Lightgrad: Lightweight diffusion probabilistic model for text-to-speech},
  author={Chen, Jie and Song, Xingchen and Peng, Zhendong and Zhang, Binbin and Pan, Fuping and Wu, Zhiyong},
  booktitle={Proceedings of the 2023 IEEE International Conference on Acoustics, Speech and Signal Processing (ICASSP)},
  pages={1--5},
  year={2023}
}

@article{pavlovic2022robust,
  title={Robust speech watermarking by a jointly trained embedder and detector using a DNN},
  author={Pavlovi{\'c}, Kosta and Kova{\v{c}}evi{\'c}, Slavko and Djurovi{\'c}, Igor and Wojciechowski, Adam},
  journal={Digital Signal Processing},
  volume={122},
  pages={103381},
  year={2022},
  publisher={Elsevier}
}

@article{de2005tutorial,
  title={A tutorial on the cross-entropy method},
  author={De Boer, Pieter-Tjerk and Kroese, Dirk P and Mannor, Shie and Rubinstein, Reuven Y},
  journal={Annals of Operations Research},
  volume={134},
  number={1},
  pages={19--67},
  year={2005},
  publisher={Springer}
}

@inproceedings{li2022dear,
  title={Dear: A novel deep learning-based approach for automated program repair},
  author={Li, Yi and Wang, Shaohua and Nguyen, Tien N},
  booktitle={Proceedings of the 44th International Conference on Software Engineering},
  pages={511--523},
  year={2022}
}

@inproceedings{oord2018parallel,
  title={Parallel wavenet: Fast high-fidelity speech synthesis},
  author={Oord, Aaron and Li, Yazhe and Babuschkin, Igor and Simonyan, Karen and Vinyals, Oriol and Kavukcuoglu, Koray and Driessche, George and Lockhart, Edward and Cobo, Luis and Stimberg, Florian and others},
  booktitle={Proceedings of the International Conference on Machine Learning},
  pages={3918--3926},
  year={2018},
  organization={PMLR}
}

@article{van2016wavenet,
  title={Wavenet: A generative model for raw audio},
  author={Van Den Oord, Aaron and Dieleman, Sander and Zen, Heiga and Simonyan, Karen and Vinyals, Oriol and Graves, Alex and Kalchbrenner, Nal and Senior, Andrew and Kavukcuoglu, Koray and others},
  journal={arXiv preprint arXiv:1609.03499},
  volume={12},
  pages={1},
  year={2016}
}

@article{wang2017tacotron,
  title={Tacotron: Towards end-to-end speech synthesis},
  author={Wang, Yuxuan and Skerry-Ryan, RJ and Stanton, Daisy and Wu, Yonghui and Weiss, Ron J and Jaitly, Navdeep and Yang, Zongheng and Xiao, Ying and Chen, Zhifeng and Bengio, Samy and others},
  journal={arXiv preprint arXiv:1703.10135},
  year={2017}
}

@inproceedings{guo2025audio,
  title={$\{$AUDIO$\}$$\{$WATERMARK$\}$: Dynamic and Harmless Watermark for Black-box Voice Dataset Copyright Protection},
  author={Guo, Hanqing and Guo, Junfeng and Chen, Bocheng and Wang, Yuanda and Chen, Xun and Huang, Heng and Yan, Qiben and Xiao, Li},
  booktitle={34th USENIX Security Symposium (USENIX Security 25)},
  pages={4601--4620},
  year={2025}
}

@article{roman2024proactive,
  title={Proactive detection of voice cloning with localized watermarking},
  author={Roman, Robin San and Fernandez, Pierre and D{\'e}fossez, Alexandre and Furon, Teddy and Tran, Tuan and Elsahar, Hady},
  journal={arXiv preprint arXiv:2401.17264},
  year={2024}
}

@article{lee2021priorgrad,
  title={Priorgrad: Improving conditional denoising diffusion models with data-dependent adaptive prior},
  author={Lee, Sang-gil and Kim, Heeseung and Shin, Chaehun and Tan, Xu and Liu, Chang and Meng, Qi and Qin, Tao and Chen, Wei and Yoon, Sungroh and Liu, Tie-Yan},
  journal={arXiv preprint arXiv:2106.06406},
  year={2021}
}

@inproceedings{panayotov2015librispeech,
  title={Librispeech: an ASR corpus based on public domain audio books},
  author={Panayotov, Vassil and Chen, Guoguo and Povey, Daniel and Khudanpur, Sanjeev},
  booktitle={Acoustics, Speech and Signal Processing (ICASSP), 2015 IEEE International Conference on},
  pages={5206--5210},
  year={2015},
  organization={IEEE}
}

@article{cox1997secure,
  title={Secure spread spectrum watermarking for multimedia},
  author={Cox, Ingemar J and Kilian, Joe and Leighton, F Thomson and Shamoon, Talal},
  journal={IEEE Transactions on Image Processing},
  volume={6},
  number={12},
  pages={1673--1687},
  year={1997},
  publisher={IEEE}
}

@inproceedings{schroter2022deepfilternet,
  title={DeepFilterNet: A low complexity speech enhancement framework for full-band audio based on deep filtering},
  author={Schroter, Hendrik and Escalante-B, Alberto N and Rosenkranz, Tobias and Maier, Andreas},
  booktitle={ICASSP 2022-2022 IEEE International Conference on Acoustics, Speech and Signal Processing (ICASSP)},
  pages={7407--7411},
  year={2022},
  organization={IEEE}
}

@article{sanli2025low,
  title={Low computational cost end-to-end speech recognition based on discrete wavelet transform and subband decoupling},
  author={Sanli, TIAN and Ta, LI and Lingxuan, YE and Shisong, WU and Qingwei, ZHAO and Pengyuan, ZHANG},
  journal={ACTA ACUSTICA},
  volume={50},
  number={2},
  pages={373--383},
  year={2025}
}

@article{he2006enhanced,
  title={An enhanced psychoacoustic model based on the discrete wavelet packet transform},
  author={He, Xing and Scordilis, Michael S},
  journal={Journal of the Franklin Institute},
  volume={343},
  number={7},
  pages={738--755},
  year={2006},
  publisher={Elsevier}
}

@article{ping2017deep,
  title={Deep voice 3: Scaling text-to-speech with convolutional sequence learning},
  author={Ping, Wei and Peng, Kainan and Gibiansky, Andrew and Arik, Sercan O and Kannan, Ajay and Narang, Sharan and Raiman, Jonathan and Miller, John},
  journal={arXiv preprint arXiv:1710.07654},
  year={2017}
}
\end{document}